\RequirePackage{fix-cm}
\documentclass[smallextended]{svjour3}       \smartqed  
\usepackage{amsmath}
\usepackage{textcomp}
\usepackage{graphicx}
\usepackage{caption}
\usepackage{todonotes}

\journalname{CEAS Space Journal}

\begin{document}

\title{Precise and robust optical beam steering for space optical instrumentation}

\author{G. Drougakis \and K. G. Mavrakis \and S. Pandey  \and G.Vasilakis \and K. Poulios \and D. G. Papazoglou \and W.~von~Klitzing}

\institute{G. Drougakis            \and
           K. G. Mavrakis 
           \and
           S.Pandey \at
              Institute of Electronic Structure and Laser, Foundation for Research and Technology-HELLAS, Heraklion 70013, Greece \at
              Department of Materials Science and Technology, University of Crete, Heraklion 70013, Greece 
            \and
             G.Vasilakis \at
              Institute of Electronic Structure and Laser, Foundation for Research and Technology-HELLAS, Heraklion 70013, Greece
            \and
            K. Poulios \at
              Institute of Electronic Structure and Laser, Foundation for Research and Technology-HELLAS, Heraklion 70013, Greece \at
              Now at: School of Physics \& Astronomy, University of Nottingham, University Park, Nottingham NG7 2RD, UK
            \and
            D. G. Papazoglou  \at
              Department of Materials Science and Technology, University of Crete, Heraklion 70013, Greece 
               \at
              Institute of Electronic Structure and Laser, Foundation for Research and Technology-HELLAS, Heraklion 70013, Greece
              \and
            Wolf von Klitzing \at
              Institute of Electronic Structure and Laser, Foundation for Research and Technology-HELLAS, Heraklion 70013, Greece 
}

\date{Received: date / Accepted: date}

\maketitle

\begin{abstract}
This approach permits much finer adjustments of the beam direction and position when compared to other beam steering techniques of the same mechanical precision. 
This results in a much increased precision, accuracy and mechanical stability. 
A precision of better than $5~\mu$rad and $5~\mu$m is demonstrated, resulting in a resolution in coupling efficiency of 0.1\%.
Together with the added flexibility of an additional beam steering element, this allows a great simplification of the design of the fiber coupler, which  normally is the most complex and sensitive element on an optical fiber breadboard.
We demonstrate a fiber to fiber coupling efficiency of more than 89.8\%, with a stability of 0.2\% in a stable temperature environment and 2\% fluctuations over a temperature range from 10\textdegree{}C to 40\textdegree{}C over a measurement time of 14 hours.
Furthermore, we do not observe any non-reversible change in the coupling efficiency after performing a series of tests over large temperature variations.
This technique finds direct application in proposed missions for quantum experiments in space \cite{Ste-Quest,Kaltenbaek2016,Carraz}, e.g.\,where laser beams are used to cool and manipulate atomic clouds.
\end{abstract}
\section{Introduction\label{intro}}
Active optics play a rapidly increasing role in space instrumentation, for example in satellite systems that have started to use LIDAR \cite{Calipso,Kallenbach:13,Cosentino2012,HeliereLWBD07}, optical communication \cite{Alphasat}, and laser ranging \cite{Ranging,Grace}.

Space-based quantum technologies and atom clocks  rely on an intricate manipulation of a number of ultra-stable laser sources leading to highly complex optical signal conditioning setups, which can be problematic for  space missions.
In some cases, this can be handled by in-fiber devices, more often, however,  the requirements exceed what can be achieved in single mode waveguide devices and optical benches handling fiber to free-space to fiber coupling are required.
Examples include cases where precise frequency shifting (acousto-optic modulators) or very high extinction ratios are needed, such as proposals for cold atom experiments in space \cite{Schuldt2015EA,Aguilera2014CAQG}.
Many components require single mode fiber to fiber coupling resulting in very stringent requirements in alignment and stability of the optical breadboard and its components.
To meet these requirements, the PHARAO cold atom clock designed for space application \cite{Pharao} needed to incorporate active stabilization of many optical components like the mirrors used to inject light into fibers.
For the LISA Pathfinder and LISA candidate systems \cite{LisaPathfinder,LISA} the optical components were attached to the breadboard using hydroxyl bonding to achieve high stability standards, albeit at the cost of extreme requirements on the manufacturing process.
Another approach used in MAIUS1 mission using a combination of different types adhesives and complex moving parts to steer the beam with high precision\cite{Duncker:14,Ressel:10}.

In this paper, we report on a novel optical beam steering technique (OBST) for fiber to free-space to fiber coupling breadboards.
We achieve robust, ultra-stable and yet extremely fine beam steering using simple optical elements, like optical wedges and plates, while at the same time reducing the complexity of the  fibre couplers.
They are mounted onto a ZERODUR breadboard using a space-certified, UV-cured adhesive.
Our technique makes it possible to adjust, with extreme precision over a finite parameters range, the beam's angle and position, whilst guaranteeing a very high degree of stability.
By separating the beam steering protocol into a coarse and a fine part OBST greatly relaxes the manufacturing requirements for the optical components, e.g.~by reducing the number of degrees of freedom required for standard elements such as the beam couplers, mirrors and beam splitters.
Overall, OBST yields highly robust, small optical breadboards for use in space missions, meeting very highest stability requirements whilst reducing the complexity of the fiber couplers in terms of manufacturing and assembly.

\section{Key features\label{sec:Key features}}
\begin{figure}[b]
\centering
\includegraphics[width=0.9\linewidth]{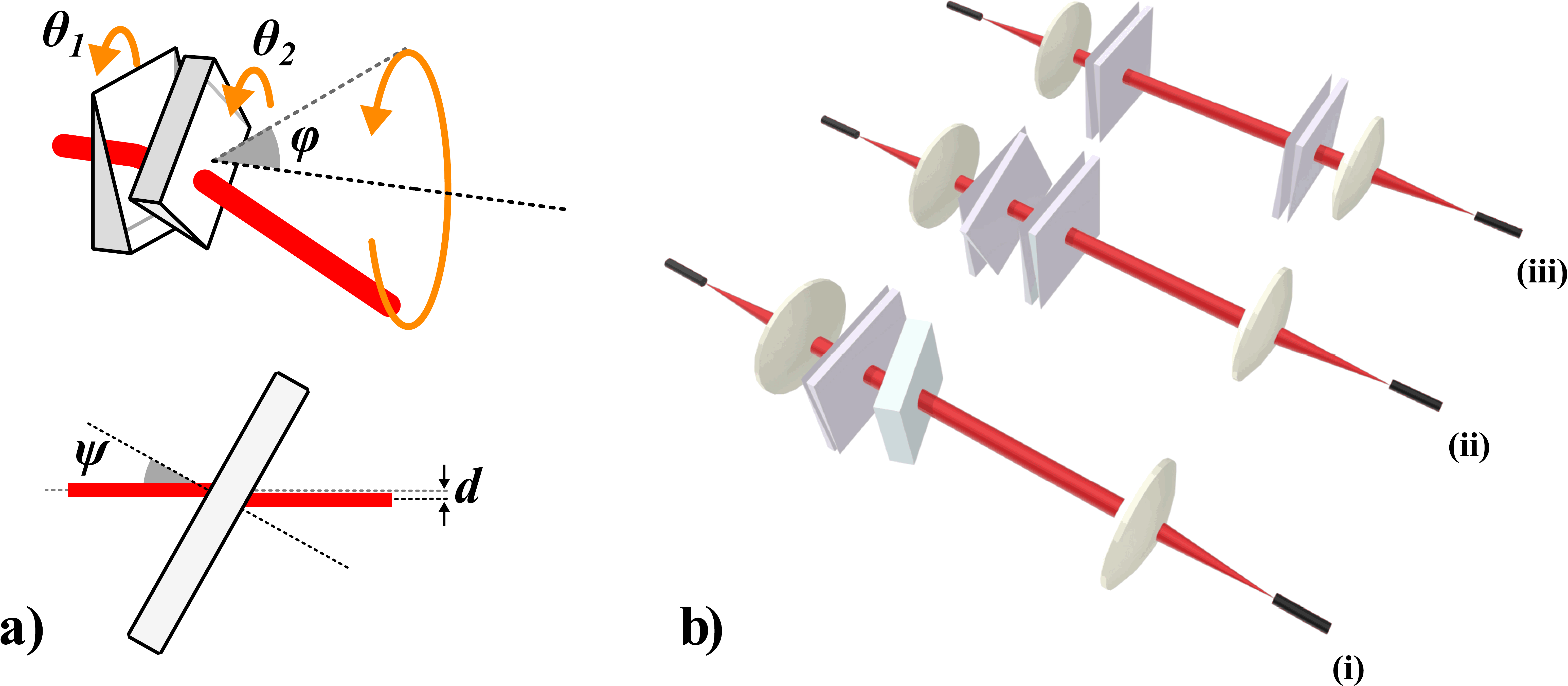}
\caption{a) Operating principle of the OBST corrective optics used in this work.b) Graphical representation of different OBST schemes: (i) one pair of wedges with a plate, (ii) 2 pairs of wedges, both at the transmitter side, (iii) 2 pairs of wedges one on the receiver one on the transmitter  \label{fig:operlatset}}
\end{figure}

OBST consists of a glass plate and/or one or two pairs of optical wedges (prisms).
The plate to displace the beam and prisms allow one to tilt it away from the optical axis it.
The operating principle of OBST is illustrated in Fig.~\ref{fig:operlatset}a.
A single prism deflects the beam by a constant angle from $\phi_1$ away from the optical axis with the direction into which the beam is deflected depending on the angle $\theta_1$ (see Fig.~\ref{fig:operlatset}a).
A second prism aligned at an angle $\theta_2$ deflects the beam again, thus either increasing the total deflection or reducing it -- depending on the relative angle between the two prisms.
Rotating  the  prism pair by an angle $\theta$  changes the angle $\theta$ into which it is deflected but keeps the total deflection angle $\phi$ constant.
A tilt $\psi$ of the glass plate leads to a parallel displacement of the beam by $d$ in the direction of the tilt.

A collimated beam has four degrees of freedom, which in the case of OBST can be controlled using either the angular degree of freedom of two distant prism pairs, or using one prism pair plus the displacement degree of freedom of the plate.
A graphical representation of various OBST configurations can be seen in Fig.~\ref{fig:operlatset}b.
Due to its relative ease of implementation the plate and wedges version of Fig.~\ref{fig:operlatset}b(i) was chosen.
In this scheme the OBST corrective optics are comprised by a pair of wedges (Thorlabs WW40530-B, 0.5 degree) and a glass plate (Eksma Optics 225-0123, 3~mm thickness). 
Our optimized system is comprised of a commercially available single mode fiber (Thorlabs PM780-HP $2w_\text{fiber}=5.3\,\mu$m mode field diameter) and molded aspheric lenses (Thorlabs 355230-B, 4.51\,mm focal length).
Using this OBST we drastically simplified the design of the optical components and the assembly process. The final performance of the OBST breadboard can be judged from Table \ref{tab:performanceTable}.

\begin{table}[t]
\centering
  \caption{Overview of the performance of the OBST prototype board 
  \label{tab:performanceTable}}
  \begin{tabular}{lrl}
    \hline
    Description & Value& Unit \\
    \hline
Coupling Efficiency & $\geq 89$ & $\%$ \\
Fluctuations of CE measured for 16h with $T=22.5$--22.8\textdegree{}C& $ \sim 0.2 $ & $ \%_\text{RMS} $ \\
\, & $ 0.8 $ & $ \%_\text{\,pp} $  \\
Fluctuations of CE measured for 24h with 10--40\textdegree{}C  ramps & $ < 2 $ & $ \%_\text{RMS} $\\
\,&$ 6 $ & $ \%_\text{ pp} $\\
Precision of the angular beam alignment& $<\,5$& $\mu$rad \\
Precision of the lateral displacement of the beam& $<\,5$& $\mu m$ \\
    \hline
  \end{tabular}
\end{table}

\section{Optical design \label{sec:Optocal design}}
To understand the basic parameters that affect the stability of a fiber to free-space to fiber coupling scheme we also performed a theoretical analysis of its performance taking into account various perturbations. 
Our analysis,  which will be presented in detail elsewhere, is based on describing the coupling efficiency, $P_{\text{tr}}$, of a freely propagating arbitrary optical field into an optical fiber, i.e. the percentage of power that can actually be transmitted, using an overlap integral:
\begin{equation}
{P_{\text{tr}}} = \frac{{{{\left| {\int  \int {{\Psi _{in}}(x,y){\Psi _0}^*(x,y)dxdy} } \right|}^2}}}{{\int  \int {{{\left| {{\Psi _{in}}(x,y)} \right|}^2}dxdy\int \int {{{\left| {{\Psi _0}(x,y)} \right|}^2}dxdy} } }},
\label{eq:CEIndx dyal}
\end{equation}
where $\Psi_{\text{in}}$ is the arbitrary distribution of the optical electric field and $\Psi_{0}$ is the fiber \emph{`eigenmode'}.
This theory was then extended to a complete fiber to free space to fiber configuration module that consists of a transmitter single mode fiber, collimating optics, receiver optics and a receiver fiber.
Given that the fibre coupler is machined from a single block of ZERODUR (with an expansion coefficient of less than $10^{-7}/\text{K}$, we  safely assume fixed distance between transmitter fiber and the collimating lens to be constant to much better than the Rayleigh length.
Based on this analysis we have evaluated the optimal parameters range for all critical elements, so that CE is above a threshold value while the sensitivity to perturbations is minimized.

A summary of our results is shown in Table \ref{tab:optcriteria} estimated for an optical link of $250~\text{mm}$ in length, using $2.5~\mu$m single mode optical fibers.
In the first row of Table \ref{tab:optcriteria} we can see that in order to achieve CE above 85\%, the receiver and collimating optics must have an effective focal length bigger than $2.1~\text{mm}$. 
The second row of Table \ref{tab:optcriteria} assesses the optimal focal length needed in order to be less sensitive to angular misalignments of the beam which can be caused by tilts of the optical components and displacements that are lateral to the optical axis.
The analysis revealed that the smaller the focal length the less sensitive to this kind of misalignment.
Finally, we examined the optimal focal length to minimize the sensitivity to the distance between the fibers along the optical axis and obtained a minimum peak at focal lengths around $4.51~\text{mm}$.
Based on these results we chose the optimal collimating and receiver optics for the proposed OBST breadboard.

\begin{figure}[t]
\centering
\includegraphics[width=0.7\linewidth]{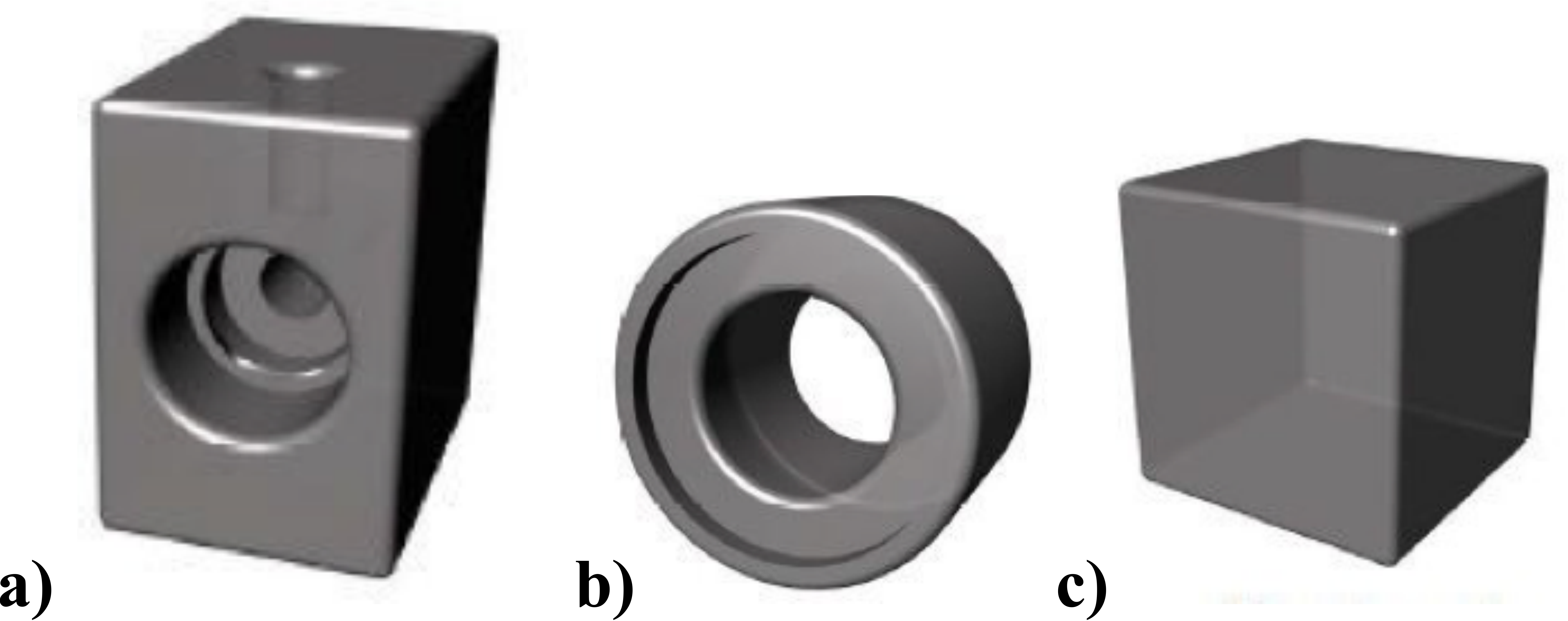}
\caption{optocal design of the ZERODUR OBST components. a) Fiber coupler (without lens), b)  Wedge holder (without wedge) and c) Mounting cube, used to secure the position of optical elements.\label{fig:components}}
\end{figure}

\begin{table}[t]
  \caption{Optimal parameters range for the focal length of the collimating lens in an optical link of a specific length (250~mm) for various criteria: i) Coupling Efficiency higher than $85\%$, ii)  Minimal sensitivity to misalignment, iii) Minimal sensitivity to variations in the distance between the fibers.\vspace{2mm}}
  \centering
  \label{tab:optcriteria}
  \begin{tabular*}{\textwidth}{c @{\extracolsep{\fill}}ccc}
    \hline
     & Criteria & Optimal parameters range & \\
    \hline
    & (i)  & $f\geq 2.1mm$ &\\
    & (ii) & $ f\xrightarrow{}0 $ & \\
    & (iii) & $f=4.51mm$ &\\
    \hline
  \end{tabular*}
\end{table}
In our experimental implementation, all the components are made of ultra-low expansion (ULE) material ZERODUR with a thermal expansion coefficient $0 \pm 0.1~\text{ppm}/\text{K}$.
The optical design of the individual components can be seen in Fig.~\ref{fig:components}.

\textbf{The fiber coupler} (seen in Fig.~\ref{fig:components}a) is a nearly monolithic device (1.5 x 1 x 1.5 cm) without any moving parts.
The design is such that the manufacturing tolerances can be achieved with standard glass machining equipment.
First, a hole is drilled to create a casing for the lens and then a second hole concentric to the first one is drilled to create a casing for the placement of the ferrule.
Finally, a ventilation hole at the top of the holder allows operation even under vacuum conditions. 

\textbf{The wedge holder} (shown in Fig.~\ref{fig:components}b) consists of a   cylinder (1.5~cm outer diameter, 1~cm thick) with the central hole has a ledge for the precise placement of the wedge.
The cubic  blocks  (1 x 1 x 1~cm, shown in Fig.~\ref{fig:components}c) serve to stabilize the wedge holders on the breadboard.

\textbf{The glass plate} is fixed by bonding its lower edge directly onto the breadboard and one side to a mounting block (1 x 1 x 1~cm as shown in Fig.~\ref{fig:components}c).
Note that there is no requirement for the edges of the plate to be exactly orthogonal.
During the alignment and adhesion process, however, the lower edge needs to be kept precisely parallel with the breadboard.

\begin{figure}[b]
\centering
\includegraphics[width=1\linewidth]{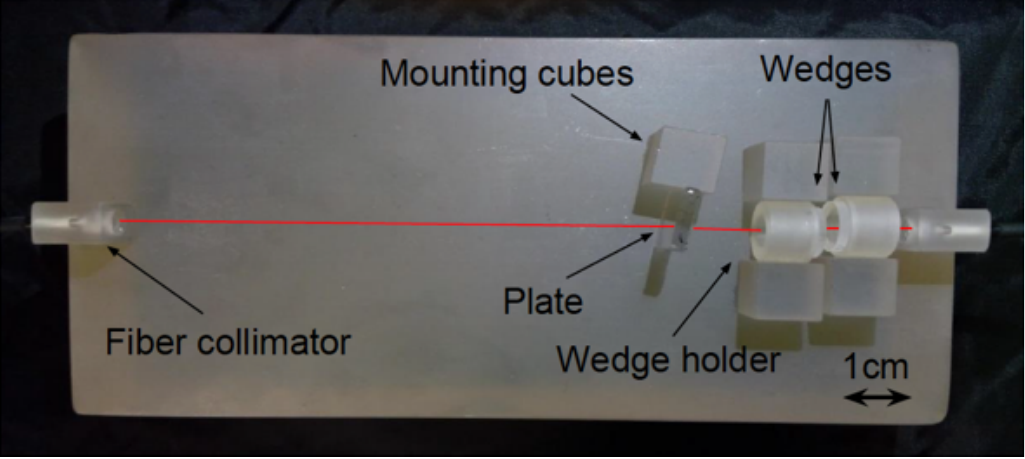}
\caption{The integrated OBST prototype breadboard. The red line represents the optical path\label{fig:breadboard}}
\end{figure}

The size of the integrated OBST breadboard (excluding optical components) shown in Fig.\,\ref{fig:breadboard} is 20 x 10 x 5~cm and is comprised of two couplers positioned at the two sides of the breadboard facing  each other, two wedge holders, a glass plate and five  mounting blocks to secure the position of the optical elements. 

\textbf{The integration} of the breadboard is performed in the following steps: First the collimators are assembled and aligned, they are then bonded onto the  breadboard after roughly pre-aligning them.Note that the angle of the wedges depends on the precision of this alignment (here 1 degree).
It should be chosen such, that the maximum expected deviation can be corrected (here, 1 degree wedge angle and 3\,mm plate thickness). 
A smaller maximum deviation will result in a smaller wedge-angle and plate thickness and thus higher resolution in alignment and a higher stability against subsequent misalignment of the plate or wedge.

\begin{figure}[ht]
\centering
\includegraphics[width=0.9\columnwidth]{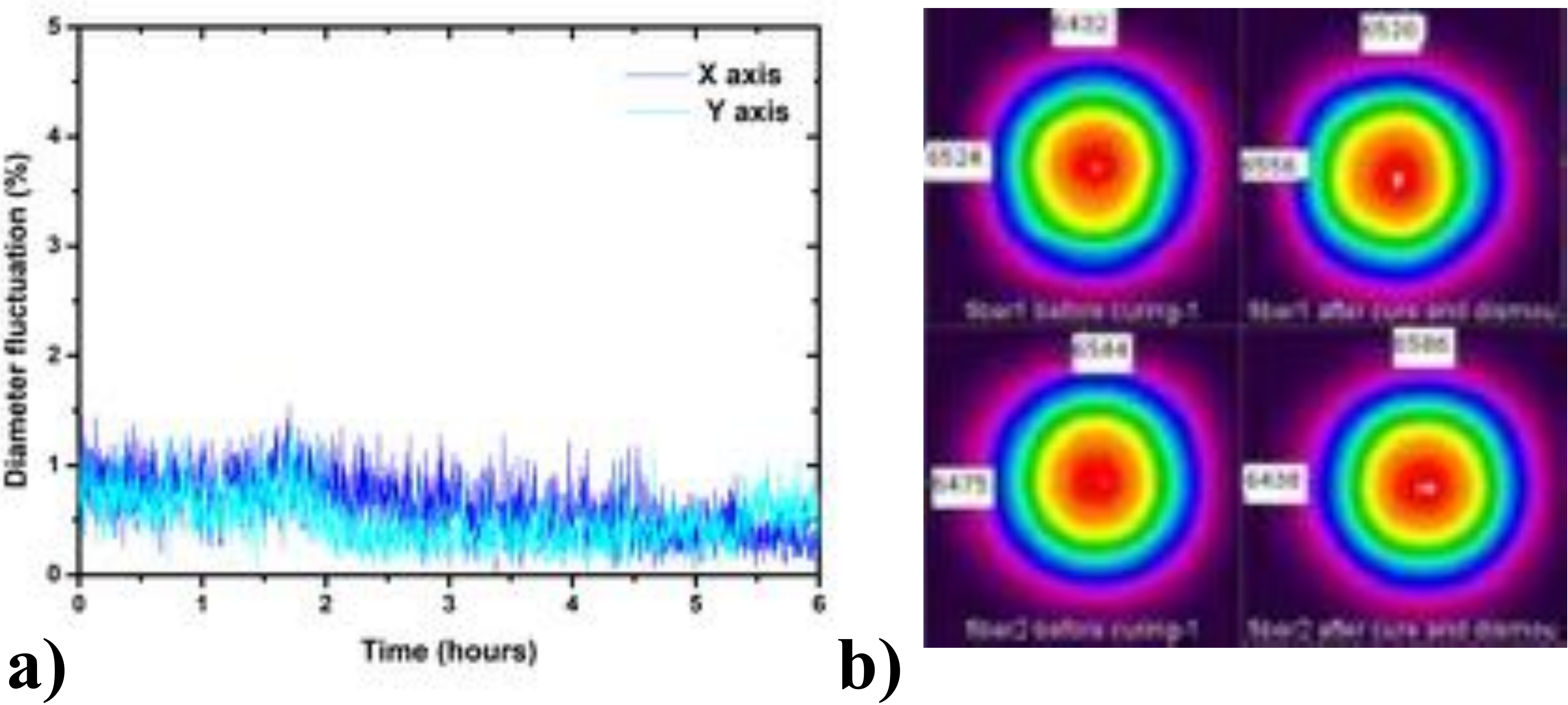}
\caption{
Beam quality and diameter during the alignment and curing process. a) Beam diameter as a function of curing time. The diameters measured were insensitive to the procedure. b) Beam profiles from two fibers fixed inside ZERODUR couplers.\label{fig:curing}}
\end{figure}

After the bonding of the coupler is complete, the light is coupled from fiber to fiber using the wedges and plate, which are then bonded in place using the mounting cubes.
In order to assemble the couplers, we first apply a thin layer of UV curing adhesive (NOA61) at the walls of the lens casing of the fiber coupler.
After placing the lens, the adhesive is cured\footnote{Note, that ZERODUR has a relatively strong absorption at 365\,nm, which increases by orders of magnitude the amount of UV light required to cure the adhesive.} using a  UV LED source (Thorlabs M365L2, 360\,mW at 365\,nm).  Next, we fix the polarization orientation and collimate the outgoing beam with the lens in place.
To achieve this the bare ferrule end of the polarization maintaining (PM) fiber is positioned using a 6 degree of freedom mount assembly consisting of a translation mount that gives the ability for xyz movement, a mirror mount with two degrees of freedom for aligning the ferrule with the coupler and a rotation mount for setting the polarization\footnote{The assembly consisted of Newport M-562 translation mount, Thorlabs KC1 mirror mount and CRM1L roation mount.}.  
We then align the polarization axis of the fiber to a polarizing beam splitter.
When this is complete, we retract the ferrule from the coupler in order to apply a layer of UV curing adhesive and proceed to the collimation.
The ferrule to lens distance is aligned using beam profiler measurements on the basis of theoretical prediction of the ideal beam parameters.
We position the ferrule to the optimum position with an accuracy better than $2~\mu \text{m}$ corresponding to an error on the coupling efficiency of less than 1\%.
The theoretical prediction for $2~\mu \text{m}$ longitudinal displacement of the ferrule is a loss of about 0.56\% which agrees well with the experimental observations. 
In Fig.~\ref{fig:curing} we show comparative measurements of the output beam diameter before and after curing.
The curing of the adhesive is performed by exposure to UV-radiation at 365~nm for 8 hours with 360~mW.
During the curing process the measured beam diameter changes by less than 2\%.
The complete coupler assembly  achieves excellent beam  characteristics with a reproducibility of 1\%  in the beam diameter. 
\begin{figure}[b]
\centering
\includegraphics[width=1\linewidth]{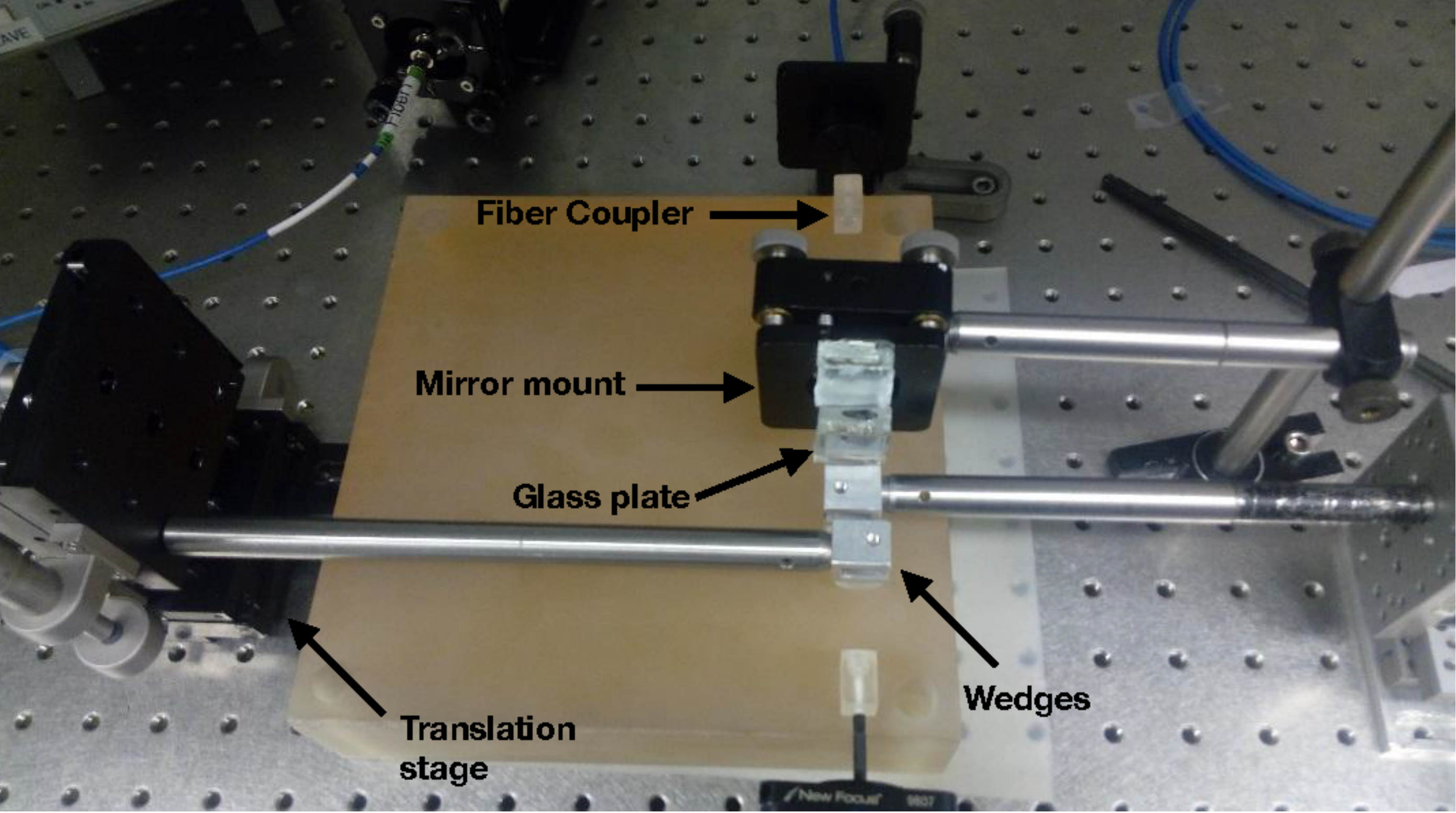}
\caption{
Setup during the alignment phase with all auxiliary alignment equipment. The metallic arms attached to the translation stages are used to roll the pair of wedges and control the pointing of the beam. The plane-parallel glass plate is attached to a standard mirror mount to control the position of the beam.\label{fig:alignment_setup}}
\end{figure}

After having collimated the couplers, we apply a drop of UV curing adhesive and place them on the breadboard, aligning them by hand judging their position by looking at the beams emerging from them.
We then cure the adhesive with a fluorescent UV lamp (Sylvania F8W/T5/BL368).
Next, the corrective optics are placed on the breadboard.
As shown in Fig.~\ref{fig:alignment_setup} the wedge holders are rolled using an extension rod which is attached to translation stages, while the plane-parallel plate is attached to a standard mirror mount.
After positioning  the corrective optics on the breadboard, we use the translation stages and the mirror mount to maximize the CE, which is continuously monitored using a Si photodiode and an oscilloscope.
Our precision of a few microradians and micrometers in tuning the beam angle and position respectively enables us to tune the CE with a precision of the order of 0.1\%.

After aligning, we fix the components in place using ZERODUR blocks.
We apply a thin layer of adhesive on the sides of the wedge and the plane-parallel plate holders and we firmly bring them in contact with the blocks.
After curing we remove all the auxiliary alignment equipment shown in Fig.\,\ref{fig:alignment_setup} and we test the breadboard.
During this process, we did not observe any significant loss in CE.

\section{Performance \label{sec:Performance}}

\begin{figure}[t]
\centering
\includegraphics[width=1\columnwidth]{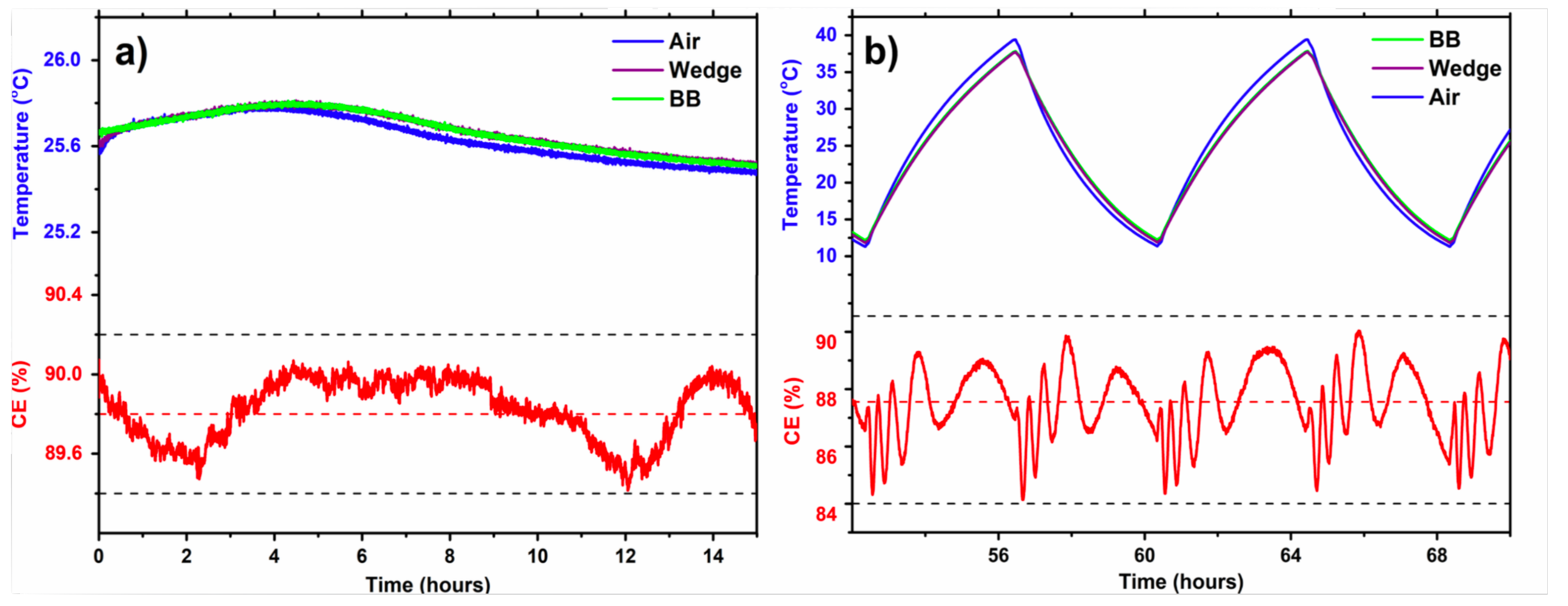}
\caption{The coupling efficiency and temperature measurements. a) Measurement with small temperature variations. Mean coupling efficiency is 89.8\% with 0.2\% RMS fluctuations. b) Two thermal cycles. Mean value of CE is 88\% with 1.7\% RMS fluctuations. The temperatures measured were the air temperature, the one at the top of one of the wedge holders and the one of the breadboard.
\label{fig:ce-temp}}
\end{figure}

In order to evaluate the performance of our breadboard in controlled temperature environment, we set up a thermal enclosure that can achieve 10-40 \textdegree{}C temperature cycling at a rate of 8\textdegree{}C/hour.
The temperature variation over time under stable conditions is shown in Fig.
\ref{fig:ce-temp}a.
Temperature is measured at three different locations using thermistors; in the air slightly above the breadboard surface (blue line), on the breadboard surface (green line) and onto the surface of one wedge holder (purple line).
As shown in the lower part of Fig.~\ref{fig:ce-temp}a under these stable temperature conditions ($25\pm 0.1$~\textdegree{}C over 15 hours), the coupling efficiency has a mean value of 89.8\% with fluctuations less than 0.8\% peak to peak and 0.2\% RMS.
Special care has been taken for the measurement of the CE to take into account the photo-diodes responsivity dependence on the temperature and position.

The performance of the breadboard during two thermal cycles can be seen in Fig.
\ref{fig:ce-temp}b.
We observe a mean value of CE=88\% with fluctuations of 1.7\% RMS.
We note two distinct Fourier components in the variations of the CE.
A fast modulation of about 2\% in amplitude and a slow variation of about 0.5-1\% in amplitude.
The slow modulation roughly follows the temperature variations and we believe it is correlated to thermal gradients in the ZERODUR base plate.
The fast component shows a sine-like fluctuation, which oscillates faster with more rapid changes in temperature.
We attribute this to interference fringes forming from the reflection off the end-facets of the optical fibers since only one of the two faces had an anti-reflection coating.
We foresee that these fringes will be reduced in strength by about one order of magnitude when all surfaces are AR-coated.
This fast component can be further reduced by angle polishing the fiber tips.
We have constructed a second OBST test breadboard with both fiber tips coated and observed a fourfold reduction (not shown here) of the amplitude of the fast fringes.

\section{Summary and outlook\label{sec:SummaryAndOutlook}}
We presented the design, manufacturing and assembly process of a ZERODUR prototype breadboard using OBST technology.
OBST is a novel beam steering technology for complex optical space applications where fiber-free space to fiber coupling is essential.
It yields highly versatile, compact, and ultra-stable optical benches without the need for any specialized alignment equipment or clean room facilities.
By separating fine from coarse alignment, OBST allows one to simplify the design of the optical components and the assembly process. 
This technology is directed towards future atom quantum experiments in space, but can also be useful where highly efficient and stable fiber coupling is needed in complex optical setups.
OBST beam steering attains a displacement resolution at the micrometer level and an angular resolution in the micro radians regime, enabling us to control the coupling efficiency to within 0.1\%.
The performance of the breadboard has been assessed in the presence of temperature fluctuations and CE of up to 90\% have been achieved with fluctuations below 0.2\%  over time scales of 15-30h.
Thermal cycling tests in a temperature range from 10-40 \textdegree{}C show fluctuations of less than 2\% RMS with no non-reversible changes due to the thermal cycling.

The thermal stability exhibited in this work is comparable with the one achieved in other approaches, like the one in MAUIS \cite{Duncker:14}, where CE between 85\% and 92\% was achieved, with the comparable advantage of ease of manufacturing and implementation design, while keeping the whole design stable by incorporating monolithic designs.
The 0.2\% RMS stability of CE in stable temperature condition is an indirect measurement of the angular stability of the beam that can be calculated to be better than $10~\mu$rad and is comparable to the one achieved using active stabilization in PHARAO laser system \cite{Pharao}.
Future work aims to assess the performance of the breadboard at the presence of vibrations and proceed to a full subsystem based on OBST that will include active optical elements such as acousto-optic modulators and beam shutters.

\section{Funding}

This work was supported by the European space agency (ESA) under contract 4000112744/14/NL/PA, the Hellenic foundation for research and innovation (HFRI) under grant agreements(4794, 4823) and the general secretariat for research and technology (GSRT).

WK would like to acknowledge the contribution of the AtomQT  COST Action CA16221.

This project has received some funding from the ATTRACT project funded by the EC under Grant Agreement 777222.

\bibliographystyle{spphys}

\end{document}